\documentclass[epj]{svjour}
\usepackage{graphics}
\newcommand\mnpbp[3]{
{\it Nucl. Phys. B {\it(Proc. Suppl.)}} {\bf #1} (#2) #3}
\newcommand{\be}{\begin{equation}}
\newcommand{\ee}{\end{equation}}
\newcommand{\ba}{\begin{eqnarray}}
\newcommand{\ea}{\end{eqnarray}}
\newcommand{\dis}{\displaystyle}
\newcommand{\mc}{\multicolumn}
\def\rs{\raisebox{1.5ex}[-1.5ex]}
\newcommand{\barr}{\begin{array}{c}}
\newcommand{\earr}{\end{array}}
\newcommand{\cF}{{\cal F}}

\def\npb#1#2#3{{\it Nucl. Phys.} {\bf B#1} (#2) #3}
\def\plb#1#2#3{{\it Phys. Lett.} {\bf B#1} (#2) #3}
\def\prd#1#2#3{{\it Phys. Rev.} {\bf D#1} (#2) #3}
\def\prl#1#2#3{{\it Phys. Rev. Lett.} {\bf #1} (#2) #3}
\def\jhep#1#2#3{{\it JHEP} {\bf #1} (#2) #3}

\def\hepph#1{hep-ph/#1}

\def\FOPTCI{$\rm FOPT_{\rm CI}$}
\def\eal{{\em et al.}}
\def\ie{{\em i.e.}} 
\def\eg{{\em e.g.}}

\def\cf{{\em c.f.}}
\def\Cl{Collaboration}

\begin{document}

\title{\begin{flushright}
\vspace{-3.2cm}
{\small\rm FTUV/01-0522 \\
 	IFIC/01-25 \\
	LAL 01-27 \\
	UG-FT-130/01 \\
	May 22, 2001 }
\end{flushright}
\vspace{1cm}
Strange Quark Mass from the Invariant Mass Distribution 
	of Cabibbo-Suppressed Tau Decays}
%
% no subtitle
% \subtitle{Do you have a subtitle?\\ If so, write it here}
\author{S. Chen\inst{1}\thanks{\emph{Now at TRIUMF, 4004 Wesbrook Mall, 
		                     Vancouver, B.C., Canada V6T 2A3}}, 
	M. Davier\inst{1}, E. G\'amiz\inst{2}, 
	A. H\"ocker\inst{1}, A. Pich\inst{3} and J. Prades\inst{2}}
%
% \offprints{}          % Insert a name or remove this line
%
% Short versions
\authorrunning{S. Chen {\em et al.}}
\titlerunning{Strange Quark Mass from the Invariant Mass Distribution 
	of Cabibbo-Suppressed Tau Decays}
\institute{Laboratoire de l'Acc\'el\'erateur Lin\'eaire, Universit\'e de
	   Paris-Sud, IN2P3-CNRS, F-91898 Orsay Cedex, France
	\and
	   Departamento de F\'{\i}sica Te\'orica y del Cosmos, 
	   Universidad de Granada, Campus de Fuente Nueva, 
	   E-18002 Granada, Spain
	\and
	   Departament de F\'{\i}sica Te\`orica, IFIC, 
	   Universitat de Val\`encia -- CSIC, 
	   Edifici d'Instituts de Paterna, Apt. Correus 22085,  
	   E-46071 Val\`encia, Spain}
%
% \date{Mai 2001}
% The correct dates will be entered by Springer
%
\abstract{
Quark mass corrections to the $\tau$ hadronic width play a significant 
role only for the strange quark, hence providing a method for
determining its mass. The experimental input is the vector plus
axial-vector strange spectral function derived from a complete study
of tau decays into strange hadronic final states performed by ALEPH.
New results on strange decay modes from other experiments are also
incorporated. 
The present analysis determines the strange quark mass at
the $M_\tau$ mass scale using moments of the spectral function.
Justified theoretical constraints are applied to the nonperturbative
components and careful attention is paid to the treatment of the
perturbative expansions of the moments which exhibit convergence 
problems. The result obtained,
$m_s(M_\tau^2)=(120\pm11_{\rm exp}\pm8_{V_{us}}\pm 19_{\rm th})~{\rm MeV}
= (120^{\,+21}_{\,-26})~{\rm MeV}$, is stable over the scale from $M_\tau$
down to about 1.4~GeV. Evolving this result to customary scales yields
$m_s(1~{\rm GeV}^2) = (160^{\,+28}_{\,-35})~{\rm MeV}$ and   
$m_s(4~{\rm GeV}^2) = (116^{\,+20}_{\,-25})~{\rm MeV} $.
%
%\PACS{
%      {PACS-key}{discribing text of that key}   \and
%      {PACS-key}{discribing text of that key}
%     } % end of PACS codes
} %end of abstract
\maketitle
\section{Introduction}

The high precision data on $\tau$ decays~\cite{TAU98,TAU00}
collected at LEP and CESR provide a powerful tool 
to investigate the dynamics of strong interaction at low 
energies and to determine basic parameters of the theory.
The QCD analysis of the nonstrange inclusive $\tau$ decay width
~\cite{BNP92,NP88,BRA89,DP92,DP92b,PIC97}
has led to accurate 
measurements~\cite{ALEPH93,CLEO95,ALEPH98,OPAL99}
of the strong coupling constant at the $\tau$ mass scale,
$\alpha_s(M_\tau^2)$,
which complements and competes in accuracy with the high precision
determination of $\alpha_s(M_Z^2)$ from measurements of the Z width at LEP.

More recently,  experimental studies of the Cabibbo-suppressed
width of the $\tau$ became available~\cite{ALEPH99,CDH99},
allowing to initiate a systematic study of the corrections
induced by the strange quark mass in the $\tau$ decay width
~\cite{BNP92,ChK93,ChK97,MA98,PP98,PP99,KM00,CKP98,KKP00}. From the 
separate measurement of the strangeness $S=0$ and $S=-1$
$\tau$ decay widths\footnote
{
  Throughout this paper, charge conjugate states are implied.
} 
it is possible to pin down the SU(3) breaking
effects and to perform a reliable determination of
the strange quark mass.

The ALEPH data~\cite{ALEPH99} have been used in previous analyses
to extract the value of $m_s(M_\tau^2)$, leading in some cases to 
different, albeit not inconsistent results~\cite{ALEPH99,PP99,KM00,KKP00}.
In this paper we present an updated common analysis~\cite{DA:00}, 
taking into account recent experimental 
information~\cite{PDG00,CLEOK99,OPAL00} in addition to the 
ALEPH data. Particular attention is given to the analysis 
of theoretical uncertainties and to the stability of the results.

Even within the relatively large statistical errors of 
the present data, the strange quark mass determination from 
$\tau$ decays has already achieved an accuracy good enough
to substantially reduce the range quoted by the Particle Data 
Group~\cite{PDG00}.

\section{Theoretical framework}

The theoretical analysis of the hadronic $\tau$ decay width
involves  the two-point correlation functions
\ba
\Pi^{\mu\nu}_{ij,V}(q) &\equiv & i {\dis \int }{\rm d}^4 x\, 
e^{iqx} \, \langle 0 | T \left\{ V_{ij}^{\mu}(x)\, 
V_{ij}^{\nu}(0)^\dagger 
\right\} |0 \rangle \, ,
\\
\Pi^{\mu\nu}_{ij,A}(q) &\equiv & i {\dis \int }{\rm d}^4 x\, 
e^{iqx} \, \langle 0 | T \left\{ A_{ij}^{\mu}(x) \, 
A_{ij}^{\nu}(0)^\dagger 
\right\} |0 \rangle \, ,
\ea
associated with the time-ordered vector,
$V_{ij}^\mu (x) \equiv  \overline q_j \gamma^\mu q_i $, 
and axial-vector,
$A_{ij}^\mu (x) \equiv  \overline q_j \gamma^\mu\gamma_5 q_i $,
colour-singlet quark currents.
The subscripts $i, j$ denote the corresponding light quark flavours 
(up, down, and strange).
The correlators admit the Lorentz decompositions
\ba
\Pi^{\mu\nu}_{ij,V/A}(q) & = &
\left( - g^{\mu\nu}\, q^2 + q^\mu q^\nu \right)
\, \Pi^{T}_{ij,V/A}(q^2) \nonumber\\
&&\hspace{0cm}
+\; q^\mu q^\nu \,
\Pi^{L}_{ij,V/A}(q^2) \, ,
\ea
where the superscript in the transverse and longitudinal components
denotes the corresponding spin, $J=1$ (T) and $J=0$ (L),
in the hadronic rest frame. 

The hadronic decay rate of the $\tau$ lepton,
\ba
\label{defrtau}
R_\tau &\equiv& \frac{\dis \Gamma \left(
\tau^-\to {\rm hadrons} \; \nu_\tau (\gamma) \right)}
{\dis \Gamma \left(
\tau^- \to e^- \, \overline{\nu}_e \, \nu_\tau (\gamma) 
\right)} \nonumber\\
& = & R_{\tau,V} + R_{\tau,A} + R_{\tau,S}
\, , 
\ea
can be expressed as an integral of the spectral functions
${\rm Im} \, \Pi^T(s)$ and ${\rm Im} \, \Pi^L(s)$ over the invariant 
mass $s$ of the  final-state hadrons, where
\ba
\label{correlators}
\Pi^J(s) &\equiv& 
	|V_{ud}|^2 \left[ \Pi_{ud,V}^J(s) + \Pi_{ud,A}^J(s)\right] \nonumber\\
	&& 
	+\; |V_{us}|^2 \left[ \Pi_{us,V}^J(s) + \Pi_{us,A}^J(s)\right] \, ,
\ea
with $|V_{ij}|$ the corresponding Cabibbo-Kobayashi-Maskawa  (CKM)
quark mixing factors.
In Eq.~(\ref{defrtau}),
$R_{\tau,V}$ and $R_{\tau,A}$ correspond to the
two terms proportional to $|V_{ud}|^2$  in (\ref{correlators})
and $R_{\tau,S}$ contains the remaining Cabibbo-suppressed 
contributions.

Using the analytic properties of $\Pi^J(s)$, one can express $R_\tau$
as a contour integral in the complex $s$-plane running counter-clockwise
around the circle $|s|=M_\tau^2$, 
\ba
\label{contour}
R_\tau &=& -\pi i \oint_{|s|=M_\tau^2} \, \frac{{\rm d}s}{s}
\left(1-{s\over M_\tau^2}\right)^3 \nonumber\\
&&\hspace{0.5cm}\times\;
\left\{ 3 
\left( 1 + {s\over M_\tau^2}\right)  D^{L+T}(s)
+ 4 \, D^L(s) \right\} \, .  
\ea
We have used integration by parts to rewrite $R_\tau$ in terms
of the logarithmic  derivatives of the relevant correlators,
\ba
D^{L+T}(s) &\equiv& -s \frac{{\rm d}}{{\rm d}s} \left[\Pi^{L+T}(s)\right]
\, ,  \\
D^{L}(s) &\equiv& \frac{\dis s}{\dis M_\tau^2} \,
\frac{\dis {\rm d}}{\dis {\rm d}s}  \left[s\, \Pi^{L}(s)\right] \, ,
\ea
which satisfy homogeneous renormalization group equations (RGE).
This  eliminates  unwanted renormalization-scheme
dependent subtraction constants  which do not contribute to
any physical observable.

For large enough $-s$, the contributions to $D^J(s)$
can be organized using the Operator Product Expansion (OPE) 
in a series of local gauge-invariant scalar operators
of increasing dimension $D=2n$,
times the appropriate inverse powers of $-s$.
This expansion is expected to be well behaved
along the complex contour $|s|=M_\tau^2$, except for the 
crossing point with the positive real axis~\cite{PQS76}.
As shown in Eq.~(\ref{contour}), the region
near the physical cut is strongly suppressed by a zero of
order three at $s=M_\tau^2$. Therefore, the uncertainties
associated with the use of the OPE near the time-like axis are
expected to be small.
Inserting the OPE in Eq.~(\ref{contour}) and evaluating the contour 
integral, one can rewrite $R_\tau$ as an expansion in inverse
powers of $M_\tau^2$~\cite{BNP92}. This leads to a rigorous
prediction for $R_\tau$ and its different flavour components, which,
in particular, allows $\alpha_s(M_\tau^2)$ to be accurately 
determined~\cite{ALEPH93,CLEO95,ALEPH98,OPAL99}.

The measurement of the invariant mass distribution of the final 
state hadrons provides additional information on the QCD dynamics. 
The moments~\cite{DP92b}
\be
\label{momdef}
R_\tau^{kl} \,\equiv\, \int_0^{M_\tau^2} \, ds \, 
\left( 1 -\frac{s}{M_\tau^2} \right)^k\, 
\left(\frac{s}{M_\tau^2}\right)^l \;
{d R_\tau\over d s} \, ,
\ee 
can be calculated theoretically in the same way as 
$R_\tau\equiv R_\tau^{00}$. 
The corresponding contour integrals take the form
\ba
\label{contourkl}
R_\tau^{kl} &=& -\pi i \oint_{|x|=1} \, \frac{{\rm d}x}{x}
\,\big\{ 3 \, \cF^{kl}_{L+T}(x) \, D^{L+T}(M_{\tau}^2 x) \nonumber\\
&&\hspace{2.5cm}
+ \; 4 \, \cF^{kl}_L(x) \, D^L(M_{\tau}^2 x) \big\} \, , 
\ea
where all kinematical factors have been absorbed into the kernels
$\cF^{kl}_{L+T}(x)$ and $\cF^{kl}_{L}(x)$.
Their explicit expressions were given in Ref.~\cite{PP99}.
Performing the contour integration, the result can be written as
\ba
\label{LAB:deltas}
R_\tau^{kl} &\equiv&  3 \left[ |V_{ud}|^2 + |V_{us}|^2 \right]
S_{\rm EW} 
\bigg\{ 1 + \delta'_{\rm EW}\, + \delta^{(0)} \\
&&\hspace{1cm}
+
{\dis \sum_{D=2,4,\cdots}} \left( {\rm cos}^2{\theta_C} \, 
\delta_{ud}^{kl (D)}+
{\rm sin}^2{\theta_C} \, \delta_{us}^{kl (D)}  \right) \bigg\} , 
\nonumber
\ea
where ${\rm sin}^2{\theta_C}\equiv |V_{us}|^2/[|V_{ud}|^2+|V_{us}|^2]$
and where we have pulled out the electroweak radiative 
corrections  $S_{\rm EW}=1.0194 \pm 0.0040$~\cite{MS88} 
and $\delta'_{\rm EW}\simeq 0.0010$~\cite{BL90}. 

The dimension-zero contribution $\delta^{(0)}$ 
is the purely perturbative correction neglecting quark masses,
which, owing to chiral symmetry, is identical for the vector and 
axial-vector parts.
The symbols $\delta_{ij}^{kl (D)} \equiv [ \delta^{kl (D)}_{ij,V}
+ \delta^{kl (D)}_{ij,A}]/2$ stand for the average of the vector and 
axial-vector contributions from dimension $D\ge 2$ operators; they
contain an implicit suppression factor $1/M_\tau^D$. 

A general analysis of the relevant $\delta^{00 (D)}_{ij,V/A}$ 
contributions was presented in Ref.~\cite{BNP92}. A 
detailed study of the perturbative piece $\delta^{(0)}$ was
later performed in Ref.~\cite{DP92}, where a resummation of
higher-order corrections induced by running effects along 
the integration contour ---denoted Contour-Improved Fixed-Order
Perturbation Theory (\FOPTCI)--- was achieved with 
renormalization-group techniques.

The leading quark-mass corrections of dimension two have been studied 
in Refs.~\cite{BNP92,ChK93,ChK97,MA98,PP98,PP99}; these contributions
are the dominant SU(3) breaking effect, which generates the wanted
sensitivity to the strange quark mass. The separate measurement 
of the Cabibbo-allowed and Cabibbo-suppressed decay widths of 
the $\tau$ allows one to pin down this SU(3) breaking effect 
through the differences~\cite{ALEPH99}
\ba
\label{diff}
\delta R_\tau^{kl} &\equiv& {R_{\tau,V+A}^{kl}\over |V_{ud}|^2} -
{R_{\tau,S}^{kl}\over |V_{us}|^2} \nonumber\\
 &=& 3 \, S_{EW}\,\sum_{D\geq 2} 
\left[ \delta^{kl\, (D)}_{ud} - \delta^{kl\, (D)}_{us} \right]
\, .
\ea
These observables vanish in the SU(3) limit, which helps to
reduce many theoretical uncertainties. 
In particular they are free of possible (flavour-independent)
instanton and/or renormalon contributions which could mimic
dimension-two corrections.
A detailed theoretical analysis of the leading $D=2$ and $D=4$
contributions to (\ref{diff}) has been given in Ref.~\cite{PP99}.
% For one-column wide figures use
\begin{figure}[t]
% Use the relevant command for your figure-insertion program
% to insert the figure file.
% For example, with the option graphics use
\resizebox{0.48\textwidth}{!}{%
  \includegraphics{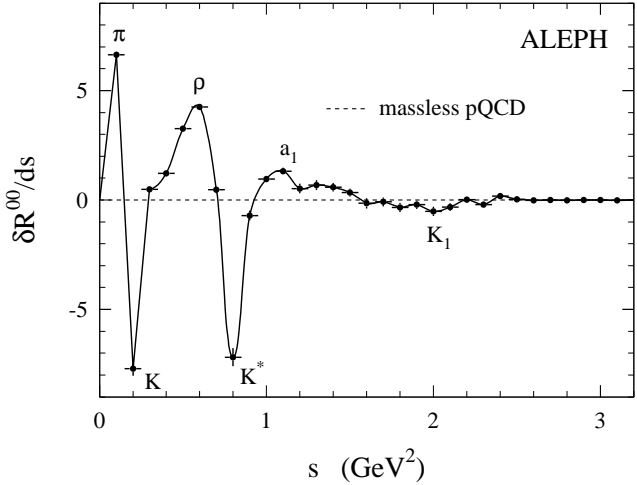}
}
% If not, use
\vspace{0.0cm}       % Give the correct figure height in cm
\caption{Integrand of Eq.~(\ref{momdef}) with (k=0, l=0) for the 
	 difference~(\ref{diff}) of the Cabibbo-corrected nonstrange 
	 and strange invariant mass spectra. The contribution 
	 from massless perturbative QCD (pQCD) vanishes.
	 To guide the eye, the solid line interpolates between 
	 the bins of constant $0.1~{\rm GeV}^2$ width.}
\label{diffmoms}       % Give a unique label
\end{figure}

\section{Experimental data}
\label{data}

ALEPH has published a comprehensive study of $\tau$ decay modes
including kaons (charged, $K^0_S$ and 
$K^0_L$)~\cite{ALEPH99,aleph_k3,aleph_ks,aleph_kl} 
with up to four hadrons in the final 
state. Such an analysis is necessary in order to separate $S=-1$
and $S=0$ modes with a $K\overline{K}$ pair.

The total branching ratio for $\tau$ decay into strange final states 
is measured to be $B_S=(2.87 \pm 0.12)\%$ corresponding to
\be
      R_{\tau,S}= 0.1610 \pm 0.0066 \, ,
\label{rtaus_aleph}
\ee
using the combined value for the electronic branching ratio,
$B_e=(17.794 \pm 0.045)\%$, obtained from the measurements of the 
leptonic branching ratios and of the $\tau$ lifetime~\cite{PDG00}, 
assuming lepton universality. Since the QCD expectation for a 
massless quark is $0.1809 \pm 0.0036$, the result~(\ref{rtaus_aleph}) 
is evidence for a massive strange quark.

%
% For tables use
\begin{table}[t]
\caption[.]{
            Measured branching ratios ($10^{-3}$) for $\tau$ decays
            into strange final states $+ \nu_\tau$. The ALEPH values are from
            Ref.~\cite{ALEPH99} and the world averages (including ALEPH) 
            from Refs.~\cite{PDG00,CLEOK99,OPAL00}. The branching ratios
            for the $(\overline{K} 4\pi)^-$ and $(\overline{K} 5\pi)^-$
            modes are estimated from the measured branching ratios for the
            $5 \pi$ and $6 \pi$ modes introducing Cabibbo and phase space
            suppression.}
\label{br}
\setlength{\tabcolsep}{0.8pc}
\begin{tabular}{lcc} 
\hline\noalign{\smallskip}
 Mode & ALEPH & World Average \\
\noalign{\smallskip}\hline\noalign{\smallskip}
  $K^-$                         & $6.96 \pm 0.29$ & $6.81 \pm 0.23$ \\
  $K^- \pi^0$                   & $4.44 \pm 0.35$ & $4.49 \pm 0.34$ \\
  $\overline{K}^0 \pi^-$        & $9.17 \pm 0.52$ & $8.78 \pm 0.38$ \\
  $K^- \pi^0 \pi^0$             & $0.56 \pm 0.25$ & $0.58 \pm 0.24$ \\
  $\overline{K}^0 \pi^- \pi^0$  & $3.27 \pm 0.51$ & $3.62 \pm 0.40$ \\
  $K^- \pi^+ \pi^-$             & $2.14 \pm 0.47$ & $2.76 \pm 0.48$ \\
  $K^- \eta$                    & $0.29 \pm 0.14$ & $0.27 \pm 0.06$ \\
  $(\overline{K} 3\pi)^-$       & $0.76 \pm 0.44$ & $0.62 \pm 0.34$ \\
  $K_1(1270)^- \rightarrow K^- \omega$        
                                & $0.67 \pm 0.21$ & $0.67 \pm 0.21$ \\
  $(\overline{K} 4\pi)^-$ (estim.)     
                                & $0.34 \pm 0.34$ & $0.34 \pm 0.34$ \\
  $(\overline{K} 5\pi)^-$ (estim.)
                                & $0.06 \pm 0.06$ & $0.06 \pm 0.06$ \\
\noalign{\smallskip}\hline\noalign{\smallskip}
  Sum                           &$28.65 \pm 1.18$ &$29.00 \pm 1.02$ \\
\noalign{\smallskip}\hline
\end{tabular}
%\vspace*{5cm}  % with the correct table height
\end{table}
The strange spectral function is derived from the  
distribution of the invariant hadronic mass.
It is dominated at low mass by the $K$ pole and the 
$K^*(892)$ resonance and at larger masses by the axial-vector 
resonances $K_1(1270)$ and $K_1(1400)$, 
decaying into $\overline{K} \pi \pi$ final states.

Fig.~\ref{diffmoms} shows the weighted integrand of the lowest 
moment $\delta R^{00}_\tau$ from the ALEPH data, as 
a function of the invariant mass-squared, and for which 
the expectation from perturbative QCD (pQCD) vanishes in the 
case of massless quarks.

The present analysis takes into account recent branching ratio results 
obtained by CLEO~\cite{CLEOK99} and OPAL~\cite{OPAL00}, thus 
improving the normalization of the individual contributions to 
the ALEPH strange spectral function.

The largest effect is found in the $\overline{K} \pi \pi$ final states.
In particular, the branching ratio for the $K^- \pi^+ \pi^-$
mode, $(2.14 \pm 0.47)\times 10^{-3}$ for ALEPH, becomes in the average
$(2.76 \pm 0.48)\times 10^{-3}$, where the inflated error takes into 
account the poor $\chi^2$ of the fit. Also, the branching ratio for
$\overline{K}^0 \pi^- \pi^0$, $(3.27 \pm 0.51)\times 10^{-3}$ for ALEPH,
becomes $(3.62 \pm 0.40)\times 10^{-3}$ in the average.

\begin{figure}[t]
% Use the relevant command for your figure-insertion program
% to insert the figure file.
% For example, with the option graphics use
\vspace{0.1cm}
\resizebox{0.48\textwidth}{!}{%
  \includegraphics{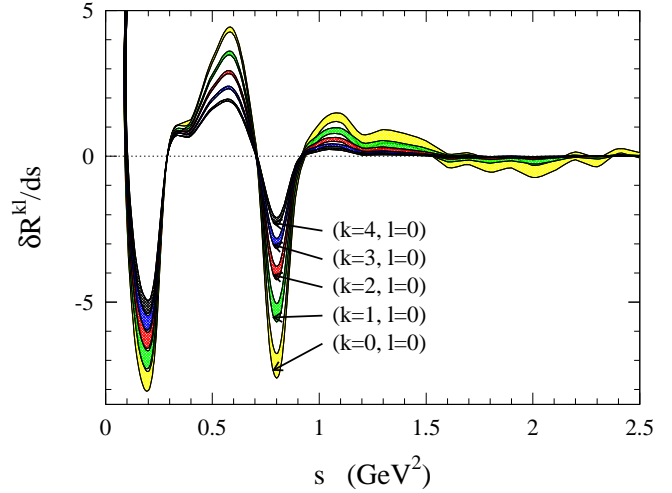}
}
% If not, use
\vspace{0.0cm}       % Give the correct figure height in cm
\caption{Integrand of Eq.~(\ref{momdef}) for the moments 
	 $k=0,\dots,4$ and $l=0$ of the difference~(\ref{diff}).
	 The widths of the interpolating bands corresponds
	 to one standard deviation.}
\label{fig_diffmom}       % Give a unique label
\end{figure}

The resulting world average for the total strange rate is found to be
\be
      R_{\tau,S}= 0.1630 \pm 0.0057 \, ,
\label{rtaus_wa}
\ee
with a $13\%$ improvement in precision. Table~\ref{br} summarizes the
different contributions to $R_{\tau,S}$.
The value for $|V_{us}|$ is taken from the Particle Data Group
unitarity fit of the CKM matrix~\cite{PDG00} yielding
$|V_{us}| = 0.2225 \pm 0.0021$.
The experimental results for the spectral moments $\delta R_\tau^{kl}$
and their correlations are given in Table~\ref{tab_dmoments}.
\begin{table}[t]
\caption[.]{
            Measured spectral moments $\delta R_\tau^{kl}$
            (top table: first error is experimental, 
	    second from $|V_{us}|$)
            and their experimental correlations
            (bottom table). }
\label{tab_dmoments}
\setlength{\tabcolsep}{3.0pc}
\begin{tabular}{cc} 
\hline\noalign{\smallskip}
 $(k,l)$& $\delta R_\tau^{kl}$ \\
\noalign{\smallskip}\hline\noalign{\smallskip}
  (0,0) & $0.374 \pm 0.118 \pm 0.062$ \\
  (1,0) & $0.398 \pm 0.065 \pm 0.042$ \\
  (2,0) & $0.399 \pm 0.044 \pm 0.031$ \\
  (3,0) & $0.396 \pm 0.034 \pm 0.024$ \\
  (4,0) & $0.395 \pm 0.028 \pm 0.020$ \\
\noalign{\smallskip}\hline
\end{tabular}

\vspace{0.3cm}
\setlength{\tabcolsep}{0.94pc}
\begin{tabular}{cccccc} 
\hline\noalign{\smallskip}
$(k,l)$ & (0,0) & (1,0) & (2,0) & (3,0) & (4,0)  \\
\noalign{\smallskip}\hline\noalign{\smallskip}
(0,0)   &  1    &  0.94 & 0.83  & 0.71  & 0.61   \\
(1,0)   & -     &  1    & 0.97  & 0.90  & 0.82   \\
(2,0)   & -     & -     &  1    & 0.98  & 0.94   \\
(3,0)   & -     & -     & -     &  1    & 0.99   \\
(4,0)   & -     & -     & -     & -     & 1      \\
\noalign{\smallskip}\hline
\end{tabular}
%\vspace*{5cm}  % with the correct table height
\end{table}
It can be observed that the central values stay rather 
constant between $k=0$ and $k=4$, while the errors decrease.
This is due to the fact that for higher $k$ values the spectral moments 
are relying more and more on the accurately measured 
$K$ and $K^*(892)$ channels. The contributions from $K\,n\pi$, $n\ge2$
modes to the moments are negligible for $k > 2$. The error 
from $|V_{us}|$ is also reduced as $R_{\tau,S}^{k0}$ decreases 
with $k$. The contributions from the various decay modes 
are visualized in Fig.~\ref{fig_diffmom}. 

\section{Phenomenological analysis}
\label{analysis}

The present analysis is performed using the general OPE framework
including the light quark masses, the $D=4$ quark mass corrections 
and $D=6$ nonperturbative contributions~\cite{BNP92}, neglecting 
higher dimensions. However, these terms are numerically 
insignificant compared to the present experimental uncertainty 
so that for all practical purposes the strange quark mass is 
obtained from the relation:
\ba
\label{mass}
m_s^2(M_\tau^2) &\simeq& \frac{M_\tau^2} {\Delta^{(2)}_{kl}(a_\tau)}
 \bigg[ \frac{\delta R_\tau^{kl}}{24 S_{EW}} \nonumber\\
&&\hspace{1cm}
+\; 2 \pi^2 \frac{\langle\delta O_4(M_\tau^2)\rangle}
		{M_\tau^4} Q_{kl}(a_\tau)\bigg] \, , 
\ea
where $\Delta^{(2)}_{kl}(a_\tau)$ and $Q_{kl}(a_\tau)$ are the
pQCD series, defined in Ref.~\cite{PP99}, associated with 
the $D=2$ and $D=4$ contributions to $\delta R_\tau^{kl}$, 
$a_\tau = \alpha_s(M_\tau^2)/ \pi$, and
\ba\label{dim4}
\langle\delta O_4(M_\tau^2)\rangle &\equiv& \langle 0| m_s\, \bar s s
  - m_d \,\bar d d | 0 \rangle (M_\tau^2) \nonumber\\
&\simeq& - (1.5 \pm 0.4) \times 10^{-3} 
~{\rm GeV}^4 \, . 
\ea

Unfortunately, the QCD series for $\Delta_{kl}^{(2)}$ turn out to
be problematic. They 
exhibit bad convergence originating from the asymptotic 
behaviour of their longitudinal components~\cite{MA98,PP98,PP99}. 
While the longitudinal part is known to third order, the $L+T$ series 
has been calculated to second order only. We estimate 
the complete third order contributions (dominated by the badly
converging, but known longitudinal part) by assuming a 
geometrical growth of the perturbative coefficients,
$c_3^{L+T} \simeq c_2^{L+T}(c_2^{L+T}/c_1^{L+T})\simeq 323$,
of the corresponding Adler function~\cite{PP99}.
%where $c_3^{L+T}$ is the unknown third order coefficient 
%occurring in the expansion
%of $\Pi^{L+T}$. Note that  $c_3^{L+T}$ = $x_3^{L+T} + 179.2$, 
%where $d_3^{L+T}$ follows the notation of Ref.~\cite{PP99}.
The value of the strong coupling constant,
$\alpha_s(M_\tau^2)=0.334\pm0.022$,
is taken from the analyses~\cite{ALEPH98,OPAL99} of the nonstrange 
$V+A$ moments $R_\tau^{kl}$. 
Correlations between this value of $\alpha_s(M_\tau^2)$ and
the moments $\delta R_{kl}^{(2)}$ are negligible since the 
uncertainty on the former is dominated by theory and the latter
by the measurement of the strange component.
Using these inputs, the pQCD series 
$\Delta_{kl}^{(2)}$ can be displayed up to third order 
using \FOPTCI~\cite{DP92}, for example:
\ba
\label{pertseries}
\Delta_{00}^{(2)}(a_\tau) &=& 0.9734 + 0.4811 + 0.3718 + 0.3371
+ \dots \nonumber \\ 
\Delta_{10}^{(2)}(a_\tau) &=& 1.0390 + 0.5576 + 0.4820 + 0.4771
+ \dots  \nonumber \\
\Delta_{20}^{(2)}(a_\tau) &=& 1.1154 + 0.6432 + 0.6082 + 0.6470
+ \dots  \nonumber \\
\Delta_{30}^{(2)}(a_\tau) &=& 1.1990 + 0.7374 + 0.7516 + 0.8507
+ \dots  \nonumber \\
\Delta_{40}^{(2)}(a_\tau) &=& 1.2880 + 0.8404 + 0.9142 + 1.0928
+ \dots \nonumber\\
\ea
The exhibited behaviour is that of asymptotic series close to 
their point of ``minimum sensitivity'' and a prescription is needed 
to evaluate the expansions and to reasonably estimate their
uncertainties. Close examination of examples
of mathematical asymptotic series suggests that a reasonable procedure 
is to truncate the expansion where the terms reach their minimum value. 
The precise prescription ---cutting at the minimum, or one order before,
including the full last term of only a fraction of it--- is somewhat 
arbitrary and this ambiguity must be reflected by a specific uncertainty
attached to the procedure. In this analysis we adopt as a rule 
keeping all terms up to (and including) the minimal one and assigning 
as a systematic uncertainty the full value of the last term retained. 
It follows that the $\Delta_{k0}^{(2)}$ series are then summed
up to third order for $k=0,1$, second order for $k=2$ and first order for
$k=3,4$. It can be remarked that the assigned truncation uncertainty is
numerically equivalent to quoting an uncertainty of 330 on $x_3^{L+T}$, \ie,
a $200\%$ error.

We disregard moments $(k,l)$ with $l\neq0$ since they
suffer from an increased dependence on the $D\ge6$
nonperturbative terms. In addition such moments carry 
little information on $m_s$.

The numerical values of the $D=4$ perturbative series $Q_{kl}$ 
are given in Table~\ref{tab:Q}. The first errors give the estimated 
theoretical uncertainties from missing higher order terms in 
the (fast converging) expansion and the second errors quote the 
changes induced	by the uncertainty in the strong coupling.

%%%%%%%%% TABLE %%%%%%%%
\begin{table}[t]
\caption{Numerical values of the relevant $D=4$ 
	perturbative expansions for $\alpha_s(M_\tau^2) = 0.334\pm0.022$. 
	The first errors give the estimated theoretical 
	uncertainties from missing higher order terms in 
	the (fast converging) series and 
	the second errors quote the changes induced
	by the uncertainty in the strong coupling.}
\label{tab:Q}
\setlength{\tabcolsep}{3.2pc}
\begin{tabular}{cc}
\hline\noalign{\smallskip}
$(k,l)$ & $Q_{kl}(a_\tau)$ \\
\noalign{\smallskip}\hline\noalign{\smallskip}
(0,0) & $1.07\pm 0.02\pm 0.01$ \\
(1,0) & $1.50\pm 0.02\pm 0.01$ \\
(2,0) & $1.92\pm 0.01\pm 0.00$ \\
(3,0) & $2.33\pm 0.01\pm 0.01$ \\
(4,0) & $2.72\pm 0.03\pm 0.02$ \\
\noalign{\smallskip}\hline
\end{tabular} 
%\vspace*{5cm}  % with the correct table height
\end{table}
%%%%%%%%%%%%%%%%%%%%%%%%%

\section{Results and discussion}
\label{results}

With the prescription given in the preceding section for the perturbative
expansion $\Delta_{kl}^{(2)}(a_\tau)$ and the data from ALEPH, 
we first derive the strange quark mass from each experimental 
$\delta R^{k0}_\tau$ moment. The results are quoted 
in Table~\ref{tab:res}. The values obtained for $m_s(M_\tau^2)$ 
are rather stable between $k=0$ and $k=4$. There is however a 
small decrease which could be of statistical nature, but
could also indicate a deterioration of the validity of the 
OPE, since larger $k$ values emphasize the low-mass contributions 
rendering the approach less inclusive. The breakdown of the error 
on $m_s(M_\tau^2)$ 
into its contributions is given in Table~\ref{tab:res}: whereas the
experimental uncertainty dominates at small $k$, the theoretical
uncertainty, which receives its main contribution from the truncation
of the perturbative series (Eq.~(\ref{pertseries})) and the renormalization 
scale, increases with $k$ and dominates for $k \geq 1$. To estimate the
latter uncertainty, the renormalization scale is varied from $0.75 M_\tau$
to $2 M_\tau$. All the theoretical errors are added in quadrature. 
Since the square of the strange quark mass is measured (\cf, \ref{mass}),
the errors on $m_s(M_\tau^2)$ are asymmetric. For reading convenience,
they are symmetrized throughout this paper, except for the final 
result.

\begin{table}[t]
  \caption[.]{
              The strange quark mass at $M_\tau$ determined from each of the
	      $\delta R^{k,l}_\tau$ experimental moments. The breakdown of
              the different sources of uncertainties 
	      corresponds to: experimental, $|V_{us}|$, 
	      $\alpha_s(M_\tau^2)$, quark condensates, truncation of
              the perturbative series $\Delta ^{(2)}_{k0}(a_\tau)$, and
              renormalization scale. The last column gives the total  
              theoretical uncertainty excluding the contribution from 
              $|V_{us}|$. Errors have been symmetrized for reading
              convenience.}
\label{tab:res}
\setlength{\tabcolsep}{0.17pc}
\begin{tabular}{ccccccccc} 
\hline\noalign{\smallskip}
     &      & 
               \mc{7}{c}{$\sigma_{m_s}$ (MeV)} \\
\rs{$(k,l)$}   & \rs{$m_s$ (MeV)} 
	&                       exp. & $|V_{us}|$ & $\alpha_s$ &
                             $\langle m_s \bar{s}s\rangle$ &
                             trunc. & R-scale & th.  \\
\noalign{\smallskip}\hline\noalign{\smallskip}
 (0,0) & 132 & 26 & 13 & 2 & 4 &  9 &  9 & 14 \\
 (1,0) & 120 & 13 &  9 & 3 & 4 & 10 & 11 & 16 \\
 (2,0) & 117 & 10 &  7 & 3 & 6 & 14 & 14 & 21 \\
 (3,0) & 117 &  9 &  8 & 2 & 8 & 19 & 16 & 27 \\
 (4,0) & 103 &  7 &  5 & 3 & 9 & 20 & 19 & 29 \\
\noalign{\smallskip}\hline
\end{tabular}
%\vspace*{5cm}  % with the correct table height
\end{table}
The fact that error contributions are given 
for the truncation and the renormalization scale, both related to the
limited number of terms in the perturbative expansion, can be considered
to be a conservative approach. In addition, one can check the quoted 
systematic uncertainty by modifying the truncation procedure: cutting off
the expansion one order less than the minimum term yields 
$m_s(M_\tau^2)$ values ranging from 143~MeV (for $k=0$) to 127~MeV 
(for $k=4$), showing a slightly better
relative stability, but deviating from the nominal results given in
Table~\ref{tab:res} by values consistent with the quoted truncation errors.
Another test of the handling of the perturbative series and its poor 
convergence is obtained by comparing the chosen contour-improved method
(\FOPTCI) to the more standard procedure of fixed order expansion
without partial resummation (FOPT). The latter method provides
$m_s(M_\tau^2)$ values in the range 121~MeV 
($k=0$) to 119~MeV ($k=4$), \ie, remarkably stable and consistent with
the nominal values within the uncertainties relevant to the treatment of
the perturbative series.

The $|V_{us}|$ uncertainty is never dominant for any value of $k$.
Other sources of systematic effects are negligible: in particular,
the uncertainties from $S_{EW}$ and higher order nonperturbative 
operators $\langle\delta O_6\rangle$ (using the estimate 
of Ref.~\cite{PP99}) lie in the range 0.3 to 0.6~MeV.

% For one-column wide figures use
\begin{figure*}[t]
% Use the relevant command for your figure-insertion program
% to insert the figure file.
% For example, with the option graphics use
%\resizebox{0.48\textwidth}{!}{%
\centerline{
\resizebox{0.75\textwidth}{!}{%
  \includegraphics{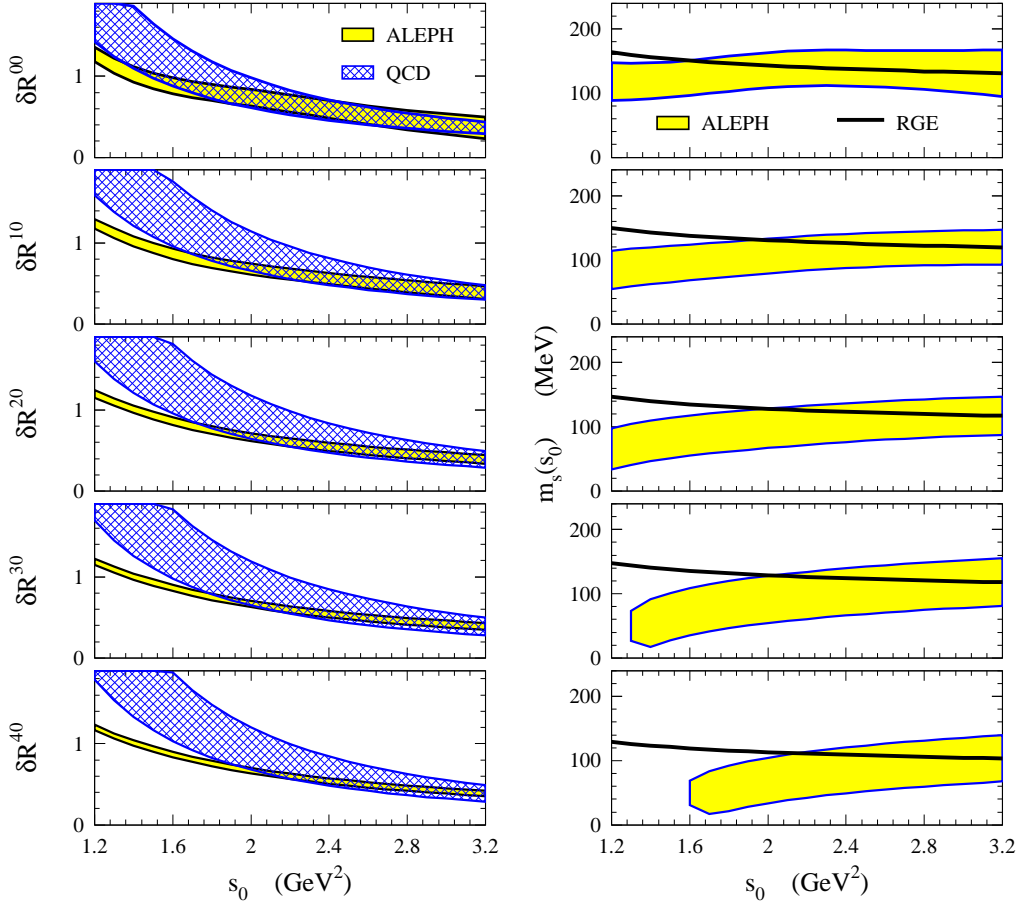}
}}
% If not, use
\vspace{-0.1cm}       % Give the correct figure height in cm
\caption{The observables $\delta R^{k0}_\tau (s_0)$ as a function of
         the ``$\tau$-mass'' squared $s_0$ confronted to the QCD predictions
         and the derived $m_s(s_0)$ compared to the QCD RGE running. The two
         bands correspond to the experimental and theoretical uncertainties.
         By construction, data and theory agree at $s_0=M_\tau^2$.}
\label{fig_running}       % Give a unique label
\end{figure*}
To optimize experimental and theoretical sensitivities, a combined 
fit is performed using several moments. The overall sensitivity 
increases up to $k=2$ while no significant improvement occurs for 
larger $k$ values and the fitted $m_s(M_\tau^2)$ 
value remains stable. Keeping only lower $k$ moments
is also justified in view of a possible breakdown of the 
OPE for higher moments and their worse perturbative convergence 
(\cf, Eq.~(\ref{pertseries})). The constrained fit of the $k=0,1,2$ 
moments takes into account the very large correlations between 
the experimental values and yields the result:
\ba
\label{resmtau}
     m_s(M_\tau^2) &\,=\,& (120 \pm 11_{\rm exp} 
			    \pm 8_{V_{us}} \pm 19_{\rm th} 
                           )~ {\rm MeV} \nonumber \\
                   &\,=\,& (120^{\,+21}_{\,-26})~ {\rm MeV} \, ,
\ea
where the theoretical error includes an additional uncertainty
of $\sigma_{\rm OPE}=12~{\rm MeV}$ which accounts for the stability 
of the result when using different moments. For reading convenience,
the errors in the first line of Eq.~(\ref{resmtau}) have been
symmetrized.

The result~(\ref{resmtau}) can be evolved to different scales using 
the four-loop RGE $\gamma$-function~\cite{ritmass}, yielding
\ba
\label{res12}
     m_s(1 ~{\rm GeV}^2) &\,=\,& (160 ^{\,+28}_{\,-35})~ {\rm MeV}\, , \\
\label{res12b}
     m_s(4 ~{\rm GeV}^2) &\,=\,& (116 ^{\,+20}_{\,-25})~ {\rm MeV}\, .
\ea

As shown in the ALEPH analysis of the nonstrange $\tau$ 
decays~\cite{ALEPH98}, one can test the validity of the QCD analysis
by simulating the physics of a hypothetical $\tau$ lepton of lower mass,
$\sqrt{s_0}$. This is obtained by replacing $M_\tau^2$ by $s_0$ 
everywhere in Eqs.~(\ref{contour}) and (\ref{momdef}), and
correcting the latter for the modified kinematic 
factor. Under the assumption of quark-hadron
duality, the evaluation of the observables as function of $s_0$
constitutes a test of the OPE approach, since the energy dependence
of the theoretical predictions is determined once the parameters 
of the theory are fixed.
The results of this exercise are given in Fig.~\ref{fig_running}, showing
the variation with $s_0$ of the first four $\delta R^{k0}_\tau$ moments
and the value for $m_s(s_0)$ derived from each moment. The bands indicate
the experimental and theoretical uncertainties. The agreement between data
and theory, perfect at $s_0=M_\tau^2$ by construction, remains acceptable
for lower $s_0$ values, down to 1.6 (2.4) GeV$^2$ for $k=0~(4)$. For the 
first three moments used in the final determination, the running observed
in data follows the RGE evolution down to about 
$2~{\rm GeV}^2$. It should be 
pointed out that the $s_0$ dependence of the theoretical prediction is 
obtained following the truncation method defined in Section~\ref{analysis}
and applied at $s_0=M_\tau^2$. If the same rule had been consistently
used at each $s_0$ point a better agreement would have been found down to
much lower $s_0$ values, at the price of introducing steps in the prediction,
corresponding to dropped terms in the expansion (according to the
prescription given in Sec.~\ref{analysis}). This observation 
provides another consistency test of the procedure.

\section{Comparison with other analyses}
\label{others}

Several analyses of the ALEPH strange spectral function have been performed
in order to extract $m_s(M_\tau^2)$. In the ALEPH paper~\cite{ALEPH99}, 
a very conservative road was followed, using only the $L+T$ part since its
perturbative expansion converges well. A price was paid in the 
experimental sensitivity, yielding rather large uncertainties on the result.
Also, the experimental moments were fitted, not only to the strange quark 
mass, but also to the values of the nonperturbative operators 
$\langle\delta O_{6,8}\rangle$.
The values obtained were in reasonable agreement with our assumption in
the present work, but unfortunately created an unlucky shift to a larger
$m_s$ value. As we have discussed here it is safe to neglect such
contributions resulting in a more constraining fit.
The value found by ALEPH is 
$m_s(M_\tau^2) = (176 ^{\,+37_{\rm exp}}_{\,-48_{\rm exp}}\,
		^{\,+24_{\rm th}}_{\,-28_{\rm th}}
                  \pm 14_{\rm meth})~{\rm MeV}$,
on the high side of the present determination for the reasons analyzed above.
The third quoted error covers uncertainties in the fit and in the 
experimental separation of $J=0,1$ states necessary to work with only the
$L+T$ part.

The method used in Ref.~\cite{PP99} is rather close to 
the present one, with the result 
$m_s(M_\tau^2) = (119 \pm 12_{\rm exp} \pm 10_{V_{us}} \pm 18_{\rm th})$ MeV.
Apart from using better experimental moments, the improvements for the 
new analysis presented here deals with a better treatment of the perturbative 
series for the different moments and an optimal use of data and theory
through a constrained fit of three moments. 

The analysis of Ref.~\cite{KKP00} 
% (which is an update of Ref.~\cite{CKP98}) 
uses the ALEPH $\delta R^{00}_\tau$ moment
and obtains $m_s(M_\tau^2) = (130 \pm 27_{\rm exp} \pm 9_{\rm th})$ MeV. 
It advocates contour-improved resummation and employs 
an effective charge as well as effective masses absorbing the 
higher perturbative terms. The central value
agrees with the corresponding result of the present analysis (given in
the first line of Table~\ref{tab:res}). The quoted theoretical uncertainty
is half as small as the one derived here, but not 
justified in the paper. Apparently no
uncertainty is included from $|V_{us}|$, which should be $\pm 13$ MeV.

Finally, the last analysis~\cite{KM00} using the ALEPH data makes use of
weight functions multiplying the correlators in Eq.~(\ref{contour}). 
These weights are tuned to improve the convergence
of the perturbative series, while suppressing the less accurate high-mass 
part of the strange spectral function. This latter feature is 
close to our use of higher moments in $k$. Since we have 
observed some deterioration of the convergence for these 
moments and correspondingly a systematic shift of the derived 
$m_s$ values, it is not completely clear to what extent this 
procedure is applicable and thus how reliable the answer is. 
Nevertheless their result is in good agreement with Eq.~(\ref{resmtau}), 
with a smaller theoretical uncertainty, apparently not including 
the effect from the arbitrary choice of the renormalization scale. 
When evolved to $M_\tau^2$, the result of Ref.~\cite{KM00} reads
$m_s(M_\tau^2)=(119\pm14_{\rm exp}\pm12_{V_{us}}\pm10_{\rm th})$~MeV.
  
Other determinations of $m_s$ have been obtained by analyses of the
divergence of the vector and axial-vector current two-point function
correlators~\cite{gasser,jamin1,chetyrkin1,becchi,dominguez1,dominguez2,jamin2,kataev,colangelo,narison1,chetyrkin2}. 
The phenomenological information on the associated scalar and pseudoscalar
spectral functions is reconstructed from phase-shift resonance analyses which
are yet incomplete over the considered mass range 
and need to be supplemented by other assumed ingredients, 
in particular the description of the continuum.

Another approach~\cite{narison3} considers the difference between
isovector and hypercharge vector current correlators as related
to the $I=1$ and $I=0$ spectral functions accessible in $e^+e^-$
annihilation into hadrons at low energy. A recent reanalysis~\cite{maltman2}
points out the possibility of large corrections from isospin breaking,
leading to significant deviations for the extracted $m_s$ value.

Finally, lattice QCD calculations of $m_s$ are available (see, 
\eg, Ref.~\cite{lattice00,MG01} for recent reviews and references 
therein), whose results are quite spread at present.
The average result reads~\cite{lattice00} 
$m_s(4~{\rm GeV}^2)=(110\pm25)~{\rm MeV}$.

\begin{figure}[t]
% Use the relevant command for your figure-insertion program
% to insert the figure file.
% For example, with the option graphics use
\centerline{
\resizebox{0.48\textwidth}{!}{%
  \includegraphics{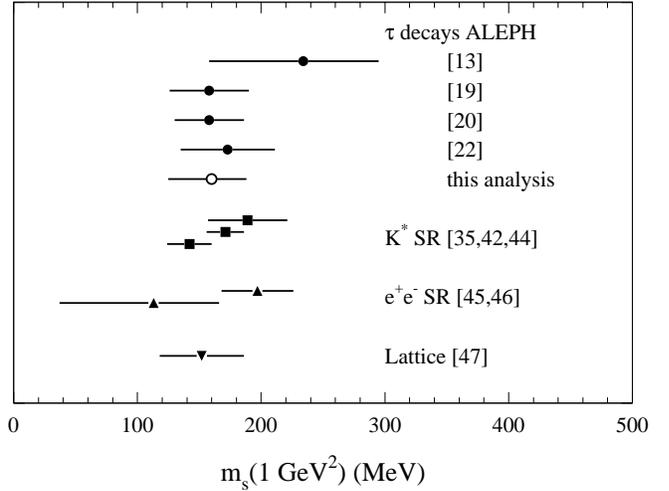}
}}
% If not, use
\vspace{0.0cm}       % Give the correct figure height in cm
\caption{The determination of $m_s(1~{\rm GeV}^2)$ in this work
         compared to the results of other analyses based on
	 the ALEPH strange spectral function and on other approaches. 
	 Details are given in the text (SR = sum rules).}
\label{mscomp}       % Give a unique label
\end{figure}

The $m_s$ determinations discussed above are compared in 
Fig.~\ref{mscomp} at the scale of $1~{\rm GeV}^2$. 

The sum of the up and down quark masses has been determined 
with Finite Energy Sum Rules~\cite{BPR95} with the result
$(m_u+m_d)(1~{\rm GeV}^2)=(12.8 \pm 2.5)~{\rm MeV}$. Using the
ratio $2m_s/(m_u+m_d)=24.4\pm1.5$, obtained within $O(p^4)$
Chiral Perturbation Theory and the large $N_C$ 
limit~\cite{Leutw98}, this result is in nice agreement 
with the present determination~(\ref{res12}).
%
% For one-column wide figures use

\section{Conclusions}

We have used the strange spectral function in $\tau$ decays measured by
ALEPH to extract the strange quark mass. The normalization has been 
improved by incorporating recent branching ratio determinations
from CLEO and LEP experiments. The stability of the $m_s$ result has
been checked by using several moments of the invariant mass distribution.
The final value is obtained from a constrained fit of 3 moments, yielding
\ba
     m_s(M_\tau^2) &\,=\,& (120 \pm 11_{\rm exp} \pm 8_{V_{us}} \pm 19_{\rm th} 
                          )~ {\rm MeV} \nonumber \\
                   &\,=\,& (120^{+21}_{-26})~ {\rm MeV}. \nonumber 
\ea
The theoretical error accounts for uncertainties, {\it (i)}, in the 
perturbative series used to $O(\alpha_s^3)$ or lower, following 
a prescription for truncating asymptotic series, {\it (ii)},
in the OPE approach, and {\it (iii)}, from smaller sources.

At the customary scales where quark masses are quoted, this result becomes 
\ba
     m_s(1~{\rm GeV}^2) &\,=\,& (160 ^{\,+28}_{\,-35})~ {\rm MeV}\,,\nonumber\\
     m_s(4~{\rm GeV}^2) &\,=\,& (116 ^{\,+20}_{\,-25})~ {\rm MeV}\,.\nonumber
\ea

The stability of the result is tested by varying the mass of a 
hypothetical $\tau$ lepton using the measured spectral function.
The analysis remains reliable within the given errors
down to a mass of about 1.4~GeV. The
procedure used for truncating the asymptotic QCD series would in fact
improve the stability down to even smaller masses, with however less
accuracy.

The uncertainty on the $m_s$ result is dominated by theory. This 
does not mean that more precise data on Cabibbo-suppressed $\tau$ 
decays are not necessary. In particular the quoted error on 
the validity of the OPE is derived from the variation of $m_s$ 
extracted from moments of increasing orders. It is not clear if the 
observed effect is of statistical nature and would not disappear with 
larger data samples. In general, more data, as expected from the 
B-factories presently in operation, will allow more checks
to be performed with a possible gain in the theoretical 
uncertainty~\cite{Gamiz}. Thus
future measurements could permit a more precise determination
of the strange quark mass along the lines presented here.

\subsection*{Acknowledgements}
{\small

J.P. would like to thank the hospitality of LAL, Univ. de Paris-Sud
at Orsay (France), where this work was initiated. E.G. is indebted
to the MECD (Spain) for a F.P.U. Fellowship. The work of E.G., A.P.
and J.P. has been supported in part by the European Union TMR
Network EURODAPHNE (Contract No. ERBFMX-CT98-0169), by MCYT Grants
No. FPA2000-1558 (E.G. and J.P.), PB97-1261 (A.P.), and by Junta 
de Andaluc\'{\i}a Grant No. FQM-101 (E.G. and J.P.).
}

%%%%%%%%%%%%%%%%%%%%%%%%

\end{document}